\documentclass[10pt, conference, letterpaper]{IEEEtran}
\usepackage[noadjust]{cite}

  \usepackage[pdftex]{graphicx}
 \usepackage{lipsum,subcaption}
\usepackage{float}
\usepackage{mathtools}  
\usepackage{amsmath}
\usepackage{amssymb,mathtools}
\usepackage{tabulary}
\usepackage{booktabs}
\usepackage{color}
\usepackage{mathptmx}
 \usepackage[export]{adjustbox}
\usepackage{blindtext}
\usepackage{multicol}
\usepackage{caption}
\usepackage{subcaption}
\usepackage{capt-of}
\usepackage[11pt]{moresize}
\usepackage{anyfontsize}

\usepackage{algorithmic,algorithm}

\usepackage{url}

\usepackage{booktabs}
\usepackage{multirow}
\makeatletter
\newcommand*{\rom}[1]{\expandafter\@slowromancap\romannumeral #1@}

\newcommand\blfootnote[1]{%
  \begingroup
  \renewcommand\thefootnote{}\footnote{#1}%
  \addtocounter{footnote}{-1}%
  \endgroup
}
\usepackage[margin=0.7in]{geometry}
\makeatother

\DeclareMathAlphabet{\mathcal}{OMS}{cmsy}{m}{n}
\SetMathAlphabet{\mathcal}{bold}{OMS}{cmsy}{b}{n}

\begin{document}
\title{A Machine Learning Approach for Power Allocation in HetNets Considering QoS\vspace{-10pt}}


\author{
    \IEEEauthorblockN{Roohollah Amiri\IEEEauthorrefmark{1}, Hani Mehrpouyan\IEEEauthorrefmark{1}, Lex Fridman \IEEEauthorrefmark{5},Ranjan K. Mallik\IEEEauthorrefmark{2}, Arumugam Nallanathan\IEEEauthorrefmark{3}, David Matolak\IEEEauthorrefmark{4}}
    \IEEEauthorblockA{\IEEEauthorrefmark{1}\textit{\small{Department of Electrical and Computer Engineering, Boise State University - Idaho, USA, \{roohollahamiri,hanimehrpouyan\}@boisestate.edu}}}
    \IEEEauthorblockA{\IEEEauthorrefmark{5}\textit{\small{Massachusetts Institute of Technology, Cambridge, MA, USA, fridman@mit.edu}}}
    \IEEEauthorblockA{\IEEEauthorrefmark{2}\textit{\small{Department of Electrical Engineering, Indian Institute of Technology - Delhi, India, rkmallik@ee.iitd.ernet.in}}}
    \IEEEauthorblockA{\IEEEauthorrefmark{3}\textit{\small{Division of Engineering, King’s College London -
London, United Kingdom, nallanathan@ieee.org}}}
    \IEEEauthorblockA{\IEEEauthorrefmark{4}\textit{\small{Department of Electrical Engineering,
University of South Carolina, Columbia, USA, matolak@cec.sc.edu}}}
}


\maketitle

\begin{abstract}
There is an increase in usage of smaller cells or femtocells to improve performance and coverage of next-generation heterogeneous wireless networks (HetNets). However, the interference caused by femtocells to neighboring cells is a limiting performance factor in dense HetNets. This interference is being managed via distributed resource allocation methods. However, as the density of the network increases so does the complexity of such resource allocation methods. Yet, unplanned deployment of femtocells requires an adaptable and self-organizing algorithm to make HetNets viable. As such, we propose to use a machine learning approach based on Q-\textit{learning} to solve the resource allocation problem in such complex networks.
By defining each base station as an agent, a cellular network is modeled as a multi-agent network. Subsequently, cooperative Q-\textit{learning} can be applied as an efficient approach to manage the resources of a multi-agent network. Furthermore, the proposed approach considers the quality of service (QoS) for each user and fairness in the network. In comparison with prior work, the proposed approach can bring more than a four-fold increase in the number of supported femtocells while using cooperative Q-\textit{learning} to reduce resource allocation overhead. \blfootnote{This research is in part supported by National Science Foundation (NSF) grant on Enhancing Access to Radio Spectrum \#1642865 and NASA ULI grant \#NNX17AJ94A.}
\end{abstract}


\IEEEpeerreviewmaketitle

\section{Introduction}
With an ever increasing density of mobile broadband users, next generation wireless networks (5G) need to support a higher density of users compared to today's networks. One approach for meeting this need is to more effectively share network resources through femtocells~\cite{art_Hani}. However, lack of guidelines for providing fairness to users and significant interference caused by unplanned deployment of femtocells are important issues that have to be resolved to make heterogeneous networks (HetNets) viable \cite{art_classic2}. In this paper reinforcement-learning (more specifically Q-\textit{learning}) as a machine learning method is used in power allocation of a dense femtocell network to maximize the sum capacity of the network while providing quality of service (QoS) and fairness to users.

\subsection{Motivation}
Ultra densification is one of the technologies to support the expected huge data traffic required of wireless networks. The idea is to use nested cells comprising small-range low-power access points called femtocells. Femtocells are connected to service providers via a broadband connection (the backhaul connection is supported via DSL or cable). As such, femtocells can be deployed by users anywhere in the cell and the overall cellular network must adapt accordingly. In the last few years, there has been concerted effort by researchers to design different algorithms to optimize the performance of femtocells within next generation wireless network, i.e., 5G. To carry the desired traffic in 5G, most of these methods have aimed for features such as reliability, fairness, and the ability to be distributive, while attempting to maintain a low complexity~\cite{art_ultra, art_macroProtect}. However, one important feature that most of these works miss is self-organization and ability to adapt to new conditions of the network. 

Reinforcement learning (RL) as a machine learning method, has been developed to optimize an unknown system by interacting with it. The nature of the RL method makes it a perfect solution for scenarios in which statistics of the system continuously change. Further, RL methods can be employed in a distributed manner to achieve even better results in many scenarios~\cite{art_complexity}. Although RL has been used in many fields, it has been just recently applied in the field of communications with specific applications in areas such as allocation problems \cite{art_reward1, art_femto, art_reward2, art_reward, art_selfOptimization, art_self_fuzzy}, energy harvesting \cite{art_harvesting}, opportunistic spectrum access \cite{art_cognitive} and other scenarios with distributed nature. With this in mind, this paper tries to apply the RL method to develop a self-organizing dense femtocell network.

\subsection{Prior Work}
The selection of a proper reward function in Q-\textit{learning} is essential because an appropriate reward function results in the desired solution to the optimization problem. In this regard, the works in~\cite{art_reward1, art_femto, art_reward2, art_reward} have proposed different reward functions to optimize power allocation between femto base stations. The works in~\cite{art_reward1, art_femto} have used independent learning while the works in~\cite{art_reward2, art_reward} have improved the prior art by using cooperative learning. The method in~\cite{art_reward1} satisfies the QoS of macro users while trying to maximize the sum capacity of the network. However, the QoS and the fairness between femto users (users served by femto base stations) are not considered. In~\cite{art_femto}, the authors try to improve the throughput of cell-edge users while keeping the fairness between the macro and femto users through a round robin approach. The work in~\cite{art_reward2} has used cooperative Q-\textit{learning} to maximize the sum capacity of the femto users while keeping the capacity of macro users near a certain threshold. Nevertheless, in both~\cite{art_femto} and~\cite{art_reward2} the QoS of femto users are not taken into consideration. Further, the reward functions in~\cite{art_reward1, art_femto, art_reward2} are not designed for a dense network. The authors in~\cite{art_reward} have used the proximity of femto base stations to a macro user as a factor in the reward function, which causes the Q-\textit{learning} method to provide a fair allocation of power between femto base stations. Their proposed reward function keeps the capacity of a macro user above a certain threshold while maximizing the sum capacity of femto users in a dense network. However, by not considering a minimum threshold for the femto users' capacity, the approaches in~\cite{art_reward1, art_reward2, art_reward, art_femto} fail to support femto users as the density of the network (and consequently interference) increases. Finally, the details of cooperation between femto base stations are not described in \cite{art_reward} and the complexity of their algorithm is not specified.

\subsection{Contribution}
In the present work, we propose a new Q-\textit{learning} approach that provides better fairness throughout the whole network. Our contributions can be categorized as follows:
\begin{itemize}
\item A new reward function is developed which satisfies the required QoS for each macro and femto user as the density of the network increases.
\item New details are provided of how to achieve cooperative Q-\textit{learning} through sharing specific rows of learning tables assigned to femto base stations to carryout power allocation between them. The proposed details clearly indicate that by using cooperative Q-learning the complexity of the machine learning approach can be significantly reduced.

\item We carry out a complexity analysis and investigation to indicate the advantage of the proposed Q-learning approach in solving resource allocation in dense HetNets.
\end{itemize}
\subsection{Organization}
 The paper is organized as follows. In Section~\ref{sec_systemModel} the system model is presented. Section~\ref{sec_problem} introduces the optimization problem and its resulting solution. Section~\ref{sec_sim} presents simulation results. Finally, Section~\ref{sec_con} concludes the paper.

\section{System Model}\label{sec_systemModel}
\begin{figure}[h]
\begin{centering}
\includegraphics[width=1\columnwidth]{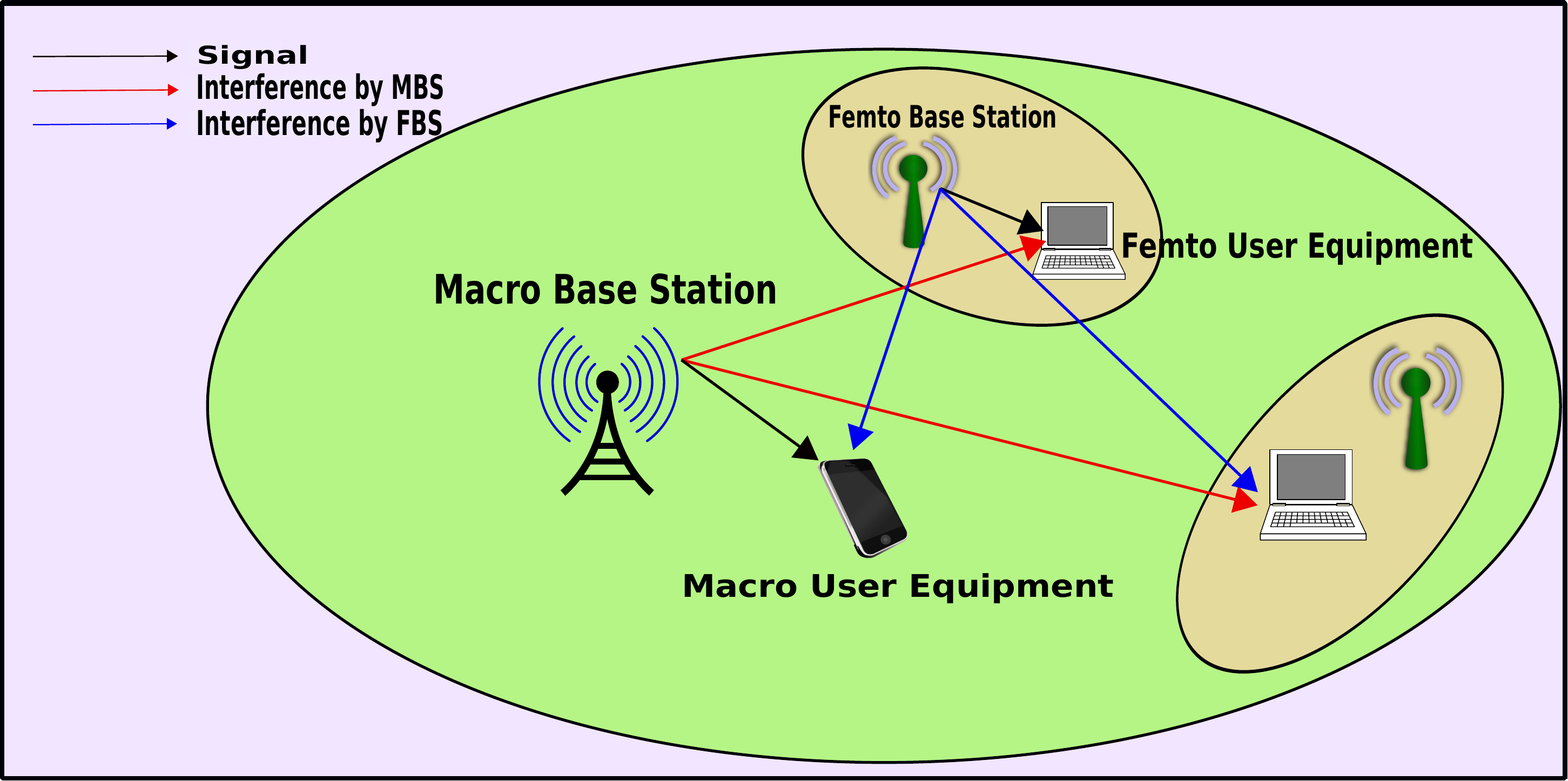}
\caption[width=.3\textwidth]{Femtocell network}
\label{fig_system}
\end{centering}
\end{figure}

In this paper we consider a single cell of a HetNet that consists of a single macro base station (MBS) and $M$ femto base stations (FBSs). Each FBS serves one user, i.e., a femto user equipment (FUE) and the MBS is assumed to serve a macro user equipment (MUE). We focus on the power allocation in the downlink of a dense HetNet, in which the density results in significant interference. All users transmit in the same spectrum, and narrowband signaling is assumed, or equivalently, results pertain to a single subcarrier of a wideband multicarrier signal. The overall network configuration is presented in Fig.~\ref{fig_system}. Note that although we consider that both the MBS and FBS server a single user, the proposed approach can easily be adapted to scenarios when more users are served.

The received signal in the downlink at the MUE receiver includes interference from the FBSs and also thermal noise. Hence, the signal-to-interference-noise-ratio (SINR) at the MUE is calculated as follows
\begin{align}\label{eq_sinr_mue}
\text{SINR}_{\text{MUE}}=\frac{P_{\text{BS}} h_{\text{BS,MUE}}}{\sum\limits_{i=1}^{M} P_i h_{{\text{FBS}_i},\text{MUE}} + \sigma^2},
\end{align}
where $P_{\text{BS}}$ is the power transmitted by MBS, $h_{\text{BS,MUE}}$ is the channel gain from the MBS to the MUE, $P_i$ is the power transmitted by the $i$th FBS, $h_{{\text{FBS}_i},\text{MUE}}$ is the channel gain from the $i$th FBS to the MUE, and $\sigma^2$ denotes the variance of the additive white Gaussian noise. 

Similarly, the SINR at the $i$th FUE is calculated as follows:
\begin{equation}\label{eq_sinr_fue}
\text{SINR}_{\text{FUE}_i}=\frac{P_i h_{\text{FBS}_i,\text{FUE}_i}}{P_{\text{BS}} h_{\text{BS,FUE}_i}+\sum\limits_{j=1,j\neq i}^{M} P_j h_{{\text{FBS}_j},\text{FUE}_i} + \sigma^2} ,
\end{equation}
where 
$h_{\text{FBS}_i,\text{FUE}_i}$ is the channel gain between the $i$th FBS and the $i$th FUE, 
$h_{\text{BS,FUE}_i}$ is the channel gain between the MBS and the $i$th FUE, $P_j$ is the power transmitted by the $j$th FBS and $h_{{\text{FBS}_j},\text{FUE}_i}$ is the channel gain between the $j$th FBS and the $i$th FUE. All channel parameters are assumed to be known by the FBS, which is consistent with prior works such as \cite{art_reward, art_self_fuzzy}. This is also practically justifiable since the channel information can be fedback to the femtocells through the bakchaul network. Finally, the normalized capacity at any user equipment is calculated as follows
\begin{equation}\label{eq_c1}
C_{\text{MUE}} = \log_2 (1+\text{SINR}_{\text{MUE}}).
\end{equation}
\begin{equation}\label{eq_c2}
C_{\text{FUE}_i} = \log_2 (1+\text{SINR}_{\text{FUE}_i}),  \; i = 1, \ldots, M.
\end{equation}

\vspace{-11pt}
\section{Problem Formulation and Proposed Solution}\label{sec_problem}
In this section, the optimization problem is defined and the Q-\textit{learning} approach to solve this problem is provided. Subsequently, the convergence of the proposed approach and cooperation between femtocells are presented.
\subsection{Optimization Problem}\label{sec_optProblem}
The goal of the optimization problem is to allocate power to the FBSs to maximize the sum capacity of the FUEs, while supporting all users (MUE and FUEs) with their required QoS. By defining $\bar{p}=\{P_1,P_2,...,P_M\}$ as the vector containing the transmit powers at the FBSs, the optimization problem can be formulated as
\begin{subequations}\label{opt_1}
\begin{align}
& \underset{\bar{p}}{\text{maximize}}
& & \sum_{k=1}^M C_{\text{FUE}_k}  \label{a}\\
& \text{subject to}
& & P_i \leq P_{max}, \; i = 1, \ldots, M\label{b}\\
&&& C_{\text{FUE}_i} \geq \tilde{q}_i , \;i = 1,...,M \label{c}\\
&&& C_{\text{MUE}} \geq \tilde{q}_{\text{MUE}}.  \label{d}
\end{align}
\end{subequations}
Here, the objective \eqref{a} is to maximize the sum capacity of the FUEs while providing the MUE with its required QoS in \eqref{d}. The first constraint, \eqref{b}, refers to the power limitation of every FBS. The terms $\tilde{q}_i$ in \eqref{c} and $\tilde{q}_{\text{MUE}}$ in \eqref{d} refer to the minimum required capacity for the FUEs and the MUE, respectively. Constraints \eqref{c} and \eqref{d} ensure that the QoS is satisfied for all users. Considering \eqref{eq_sinr_fue}, \eqref{eq_c2}, and \eqref{opt_1}, it can be concluded that the optimization in \eqref{opt_1} is a non-convex problem for dense HetNets. This follows from the SINR expression in \eqref{eq_sinr_fue} and the objective function \eqref{opt_1}. More specifically, the interference term due to the neighboring femtocells in the denominator of \eqref{eq_sinr_fue}, ensures that the optimization problem in \eqref{a} is not convex. This interference term may be ignored in low density networks but cannot be ignored in dense HetNets consisting of a large number of femtocells. 

In the next section, a Q-\textit{learning} based approach to solve this problem is proposed.

\subsection{Reinforcement Learning}\label{sec_RLTheory}
\begin{figure}[h]
\begin{center}
\includegraphics[width=1\columnwidth]{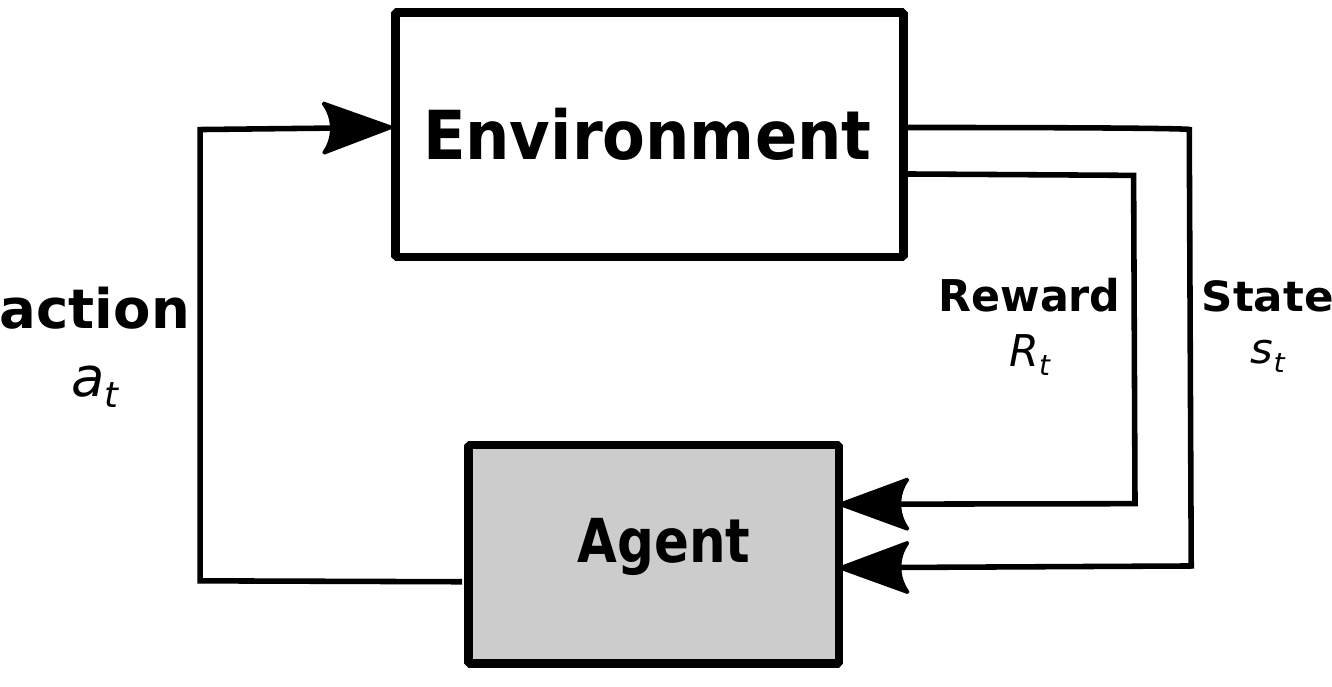}
\caption[width=.3\textwidth]{Reinforcement Learning, Agent and Environment.}
\label{fig_RL}
\end{center}
\end{figure}

The RL problem consists of an environment and a single or multiple agents which based on a chosen policy take actions to interact with the environment. After each interaction, the agent receives a feedback (reward) from the environment and updates its state. An agent can be any intelligent member of the problem, for example in a cellular network it could be an FBS. The goal of this approach is to maximize the cumulative received rewards during an infinite number of interactions. Fig.~\ref{fig_RL} shows the RL procedure. Most of the RL problems can be considered as Markov Decision Processes (MDPs). 
\subsection{Proposed Q-\textit{learning} Approach}\label{sec_RLProblem}
Q-\textit{learning} is a model-free RL method that attacks MDP problems with dynamic programming~\cite{Watkins1992}. Q-\textit{learning} can be considered as a function approximator in which the values of the approximator, Q, depend on the state ($x_t$) and action ($a_t$) at time step $t$. The dynamic programming equation for computing a function approximator Q (also known as Bellman equation) is as follows
\begin{equation} \label{eq_bellman}
Q(x_t, a_t) = \max_{a}(E[R_t + \gamma Q(x_{t+1}, a)]),
\end{equation}
where $E$ denotes the expectation operator and $R_t$ is the received reward at time step $t$ and $0 \leq \gamma \leq 1$ is the discount factor. Eq. \eqref{eq_bellman} has a unique strictly concave solution and the solution is approached by limit as $t \rightarrow \infty$ by iterations~\cite{book_DP}.

The novelty of Q-\textit{learning} is attributed to the use of \textit{temporal-difference} (TD) to approximate a Q-function~\cite{book_sutton, phdthesis_watkins}. The simplest form of \textit{one-step} Q-\textit{learning} approach, is defined by
\begin{equation} \label{eq-qlearning}
Q(x_t, a_t) \leftarrow (1-\alpha)Q(x_t, a_t) + \alpha \max_{a}(R_t + \gamma Q(x_{t+1}, a)),
\end{equation}
where $\alpha$ is the learning rate. Algorithm 1 specifies the Q-\textit{learning} in procedural form \cite{book_sutton}.
\begin{algorithm}
\renewcommand\thealgorithm{}
\caption{\textbf{1} Q-Learning algorithm}\label{alg_1}
\begin{algorithmic}[1]
\STATE  Initialize $Q(x_t,a_t)$ arbitrarily
\FORALL{episodes}
\STATE Initialize $x_t$
\FORALL{steps of episode}
\STATE Choose $a_t$ from set of actions
\STATE Take action $a_t$, observe $R_t$, $x_{t+1}$
\STATE \footnotesize\begingroup
   $Q(x_t, a_t) \leftarrow (1-\alpha)Q(x_t, a_t) + \alpha \max_{a}(R_t + \gamma Q(x_{t+1}, a))$
\endgroup
\normalsize
\STATE $x_t \leftarrow x_{t+1}$;
\ENDFOR
\ENDFOR
\end{algorithmic}
\addtocounter{algorithm}{-1}
\end{algorithm}

In the context of a femtocell network, FBS acts as an agent in the Q-\textit{learning} algorithm, which means each FBS runs Algorithm 1, separately. The Q-\textit{learning} approach consists of three main parts as follows
\subsubsection{\textbf{Actions}}
Each FBS chooses its transmit power from a set $A=\left\lbrace a_1, a_2, ... , a_{N_{power}} \right\rbrace$, which covers the space between $P_{min}$ and $P_{max}$. In general, there is no particular information from environment, so the FBS chooses actions with the same probability. Therefore, equal step sizes are chosen between $P_{min}$ and $P_{max}$.
\subsubsection{\textbf{States}}
States are chosen based on the vicinity of the FBS to the MBS and the MUE. In order to specify the state of an FBS, we define two parameters for each FBS:
\begin{itemize}
\item {$D_{\text{MBS}} \in \left\lbrace 0,1,2,...,N_1 \right\rbrace$}: The value of $D_{\text{MBS}}$ defines the location of an FBS compared to $N_1$ rings centered on the MBS. The radius of layers are $d_{\text{BS}1}$, $d_{\text{BS}2}$, ... , $d_{\text{BS}_{N_1}}$.
\item {$D_{\text{MUE}} \in \left\lbrace 0,1,2,...,N_2 \right\rbrace$}: The value of $D_{\text{MUE}}$ defines the location of an FBS compared to $N_2$ rings centered on the MUE. The radius of layers are $d_{\text{FBS}1}$, $d_{\text{FBS}2}$, ... , $d_{\text{FBS}_{N_2}}$.
\end{itemize}
By considering the above definitions, the state of the FBS $i$ at time step $t$ is defined as $s_t^i \in \left\lbrace D_{\text{MBS}}, D_{\text{MUE}}\right\rbrace$. Each FBS, constructs a table for itself, which comprises all possible states as its rows and actions as its columns called a Q-table. By the state definition, in the proposed model, the FBS state remains constant as long as its location is fixed. This feature brings an advantage in sharing Q-tables between the FBSs with the same state, where only a single row of each FBS's Q-table needs to be shared.

\subsubsection{\textbf{Reward}}
The definition of the reward function is essential because it targets the objective of the Q-\textit{learning} method. According to the optimization problem in \eqref{opt_1}, the goal of the optimization process is to maximize the sum capacity of femto users in the network while maintaining QoS for each one of them. In order to translate this objective to a reward function, the following points are taken into account:
\begin{itemize}
\item The objective of the optimization problem is to maximize the capacity of the network, so a higher capacity for FUE or MUE should result in a higher reward.
\item To satisfy the QoS requirements of users, capacity deviation of users from their required QoS ($\tilde{q}_i$ or $\tilde{q}_{\text{MUE}}$) should decrease the reward.
\end{itemize} 
By considering the above points, the proposed reward function (RF) for the i$th$ FBS at time step $t$ is defined as
{\footnotesize
\begin{flalign}\label{RF1}
  R_t^i =
    \underbrace{\beta_i}_{(d)} \underbrace{C_{\text{FUE}_{i,t}} C^2_{\text{MUE}_t}}_{(a)}- \underbrace{\frac{1}{\beta_i}}_{(e)} \underbrace{\left( C_{\text{MUE}_t}-\tilde{q}_{\text{MUE}}\right)^2}_{(b)} - \underbrace{\left( C_{\text{FUE}_{i,t}}-\tilde{q}_i \right)^2}_{(c)},
\end{flalign}
}


\noindent
which is derived based on the above points. In \eqref{RF1}, $C_{\text{FUE}_{i,t}}$ and $C_{\text{MUE}_t}$ are the capacities of the $i$th FUE and the MUE at time step $t$, respectively. According to \eqref{RF1}, the reward function comprises three main terms \textit{(a), (b),  and (c)}. The first term (\textit{a}) implies that a higher capacity for the $ith$ FUE or the MUE results in a higher reward. In the same term, the capacity of the MUE is squared. The power assigned to $C_{\text{MUE}_t}$ in (\textit{a}), is supported by our simulation in Section~IV and to also give a higher priority to MUE with respect to the FUE by allocating a higher reward to its capacity value. The terms (\textit{b}) and (\textit{c}) are deviations of the $ith$ FUE and the MUE from their required threshold. Hence, terms (\textit{b}) and (\textit{c}) are reduced from term \textit{a} to decrease the reward. Terms (\textit{d}) and (\textit{e}) provide fairness to the algorithm. $\beta_i$ (term \textit{d}) is defined as the distance of the $ith$ FBS to the MUE normalized by $d_{th}$. $d_{th}$ is a constant distance, which indicates whether the FBS is in the vicinity of the MUE or not. For example, if the distance of the $ith$ FBS and the MUE is less than $d_{th}$, the interference of the $ith$ FBS affects the MUE more than any other FBS with distance more than $d_{th}$. Then the $ith$ FBS should be given less reward, which means reducing term (\textit{a}) by multiplying it by $\beta_i$ (or (\textit{d})) and increasing term b by multiplying it by the inverse of $\beta_i$ (or (\textit{e})). 

\subsection{Convergence}

According to~\cite{Watkins1992}, in Algorithm 1, if all actions are repeatedly sampled in all states, \eqref{eq-qlearning} will be updated until the value of $Q$ converges to the optimal value ($Q^{*}$) with probability $1$. In practice, the number of updates is limited. Hence, the final value of $Q$ may be suboptimal. Q-\textit{learning} itself is a greedy policy since it finds the action which derives the maximum Q-\textit{value} on each iteration. Greedy policies have the disadvantage of being vulnerable to environmental changes, and they can be trapped or biased in a limited search area which causes the algorithm to converge slower. One reasonable solution is to act greedy with probability $1-\epsilon$ (exploiting) and act randomly with probability $\epsilon$ (exploring). Different values for $\epsilon$ provide a trade-off between exploitation and exploration. Algorithms that try to explore and exploit fairly are called SARSA or $\epsilon$-greedy~\cite{book_sutton}. In~\cite{book_sutton} it is shown that in a limited number of iterations, the $\epsilon$-greedy policy has a faster convergence rate and closer final value to the optimal one, compared to the greedy policy. As such, the $\epsilon$-greedy policy has been used in the rest of this paper. Further, our investigations show that $\epsilon$ values of 0.1 or 0.01 provide a reasonable trade off between exploitation and exploration.

\subsection{Cooperative Q-\textit{Learning}} \label{sec_cooperation}

The time complexity of an RL algorithm depends on three main factors: the state space size, the structure of states, and the primary knowledge of the agents~\cite{art_expertness,art_complexity}. If priori knowledge is not available to an agent or if environment changes and the agent has to adapt, the search time can be excessive~\cite{art_complexity}. Considering the above, decreasing the effect of state space size on learning rate and providing agents with priori knowledge has been a subject of significant research~\cite{art_complexity, art_expertness, art_prey, art_marl_survey}. 

One approach to deal with this problem is by transferring information from one agent to another instead of expecting agents to discover all the necessary information. In fact, by using a multi agent RL network (MARL), agents can communicate and share their experiences with each other, and learn from one another~\cite{art_complexity}. The reason that cooperation can reduce the search time for RL algorithms can be attributed to the different information that the agents can gather regarding their experiences in the network. By sharing information between experienced and new agents, a priori knowledge is provided for new agents to reduce their search time. It is worth mentioning that even in a MARL network that consists of a large number of new agents, cooperation and information sharing among these agents can reduce search time for the optimum power allocation solution~\cite{art_complexity}. Another reason why cooperation enhances search speed is the inherent parallelism in cooperation between agents~\cite{art_complexity, art_marl_survey}. In other words, by sharing their information, the agents search different choices in parallel which decreases the search time greatly. In~\cite{art_complexity} it is shown that by intelligent sharing of information between agents, search time can be executed as a linear function of the state-space size.

Sharing Q-values in MARL networks for resource allocation and management is still an open research problem. The main challenge lies in the fact that the agents must be able to acquire the required information from the shared Q-values~\cite{art_expertness}. As a result, in a large MARL network it is not yet clear what Q-values must be shared among the agents to reduce search time and reach the optimum power allocation solution. Moreover, cooperation comes at the cost of communication. Agents can share their information to help each other to learn faster while adding more overhead to the network by passing on their Q-values through the backhaul network. Nevertheless, it is important to note that these Q-values can be significantly quantized to reduce this overhead.

In a femtocell network, each FBS gathers information regarding the network. The nature of this information for each FBS may be different and directly related to its active time in the network. Accordingly, we propose a cooperative Q-learning approach where the Q-tables of FBSs that are in the same state, i.e., the FBSs that are located in the same vicinity (rings) relative to the MBS and the MUE, are shared with one another. The latter is proposed since our results show that only the FBSs with the same state have useful information for one another. Moreover, the proposed approach reduces the communication overhead among the FBSs.

Accordingly, the proposed method for the femtocell network consists of two modes: individual learning and cooperative learning. The individual learning starts by initializing the Q-values of a small subset of FBSs, e.g., four, to zero. These FBSs execute the proposed RL algorithm independently. After convergence, new agents are added to the problem one by one and cooperative Q-\textit{learning} takes place. In this mode, the MARL network consists of experienced FBSs and one new FBS. The new FBS takes its priori knowledge from the FBSs with the same state and all FBSs execute the RL algorithm. The FBSs with the same state share their Q-tables (just one row) after each iteration. To form a new Q-table from the shared Q-tables, we have used the method in \cite{art_prey}, where the shared Q-tables are averaged over. Although this method is suboptimal~\cite{art_expertness} and to perform accurate sharing, a weighted averaging between Q-tables should be used, we have chosen to select the simple averaging method to achieve a lower overall complexity. 
 
\section{Simulation Results} \label{sec_sim}
In this section the simulation setup is detailed and then the results of the simulations are presented.
\subsection{Simulation Setup}\label{sec_setup}

\begin{figure}[h]
\begin{centering}
\includegraphics[width=1\columnwidth]{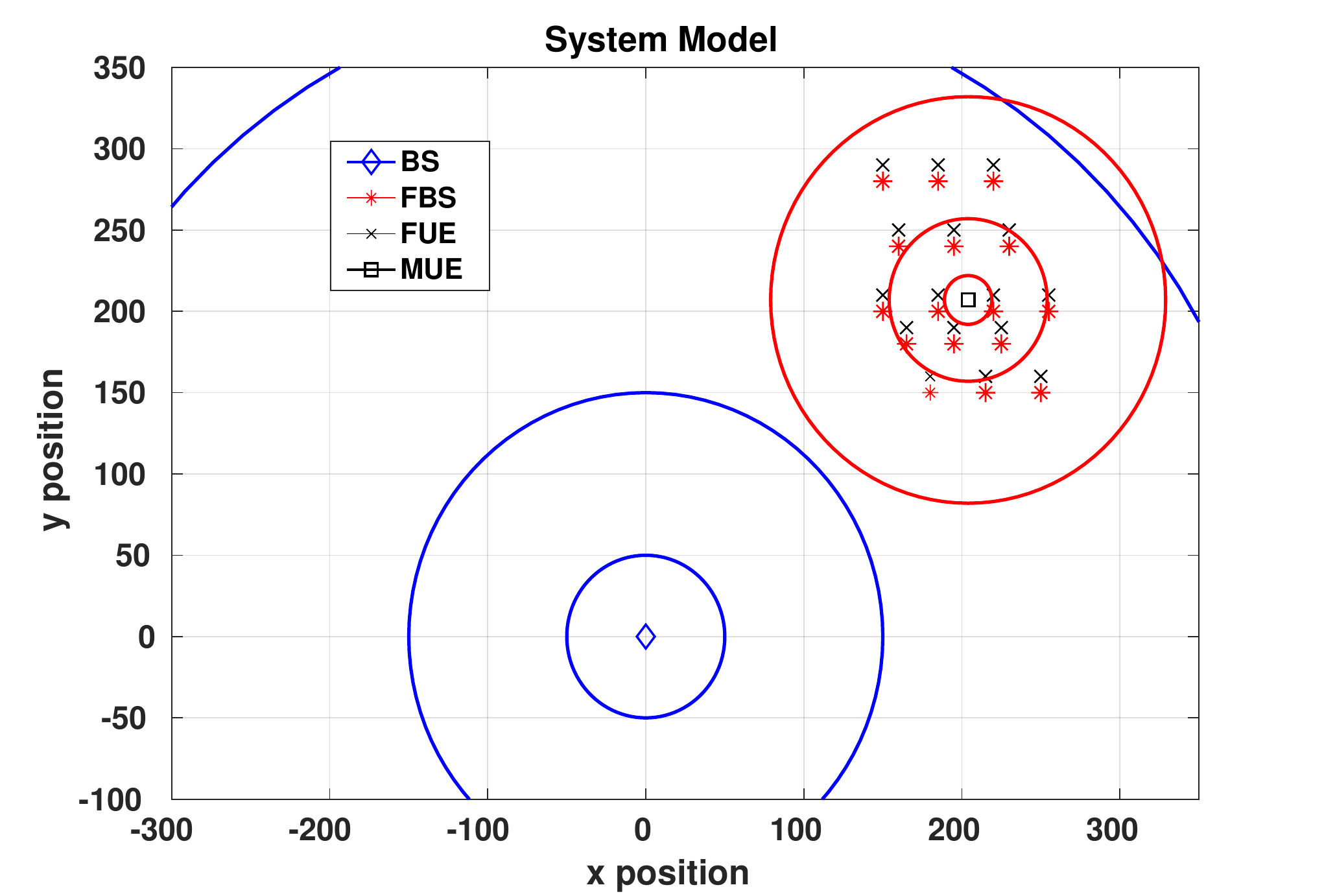}
\caption[width=.3\textwidth]{Locations of MBS, FBSs, MUE and FUEs.}
\label{fig_sim}
\end{centering}
\end{figure}

A femtocell network is simulated with a single MBS, one MUE, and M number of FBSs, where each FBS supports one FUE, see Fig.~\ref{fig_sim}. 

To simulate a residential neighborhood, the FBSs are located $35$ m apart from each other. Each FUE is located in a $10$ m radius from its serving FBS. To simulate a high interference scenario in a dense network, the MUE is located among $15$ number of FBSs. The locations of the MBS, FBSs, and MUE are shown in Fig.~\ref{fig_sim}. In these simulations the number of layers around the MBS and the MUE are assumed to be three ($N_1=N_2=3$). Although, as the density increases, more rings with smaller diameters can be used to more clearly distinguish between the FBSs. The blue and red rings indicate the states of the FBSs with respect to the MBS and the MUE, respectively.

It is assumed that the FBS and the MBS are both operating over the same channel bandwidth at $f=2.4$ GHz. The path loss model of the link between the MBS and the MUE, and the one between the FBS and its serving FUE is given by
\begin{equation}\label{path_1}
PL = PL_0 + 10 n \log_{10}(d/d_0),
\end{equation}
where $PL_0$ is the constant path loss value, and $n$ is the path loss exponent. The parameters of the model are set to: $d_0$ = $5$ m, $PL_0$ = $62.3$ dB and $n$ = $4$~\cite{book_green}, as an example of a model for a residential area. The path loss of the link between each FBS and the MUE, and the link between each FBS and the FUE of other FBSs are modeled using an empirical indoor-to-outdoor model suitable for femtocells from~\cite{pathloss}. Using (6), (7), and Table~\rom{1} from~\cite{pathloss}, the path loss can be written as
\begin{equation}
PL=PL_i+PL_0,
\end{equation}
\begin{equation}\label{path_freq}
PL_i=-1.8f^2+10.6f+6.1,
\end{equation}
\begin{equation}
PL_0=62.3+32\log_{10}(d/5),
\end{equation}
where $f$ denotes the operating frequency in GHz. The remaining parameters are given in Table~\ref{table_1}.

\begin{table}[h]
\centering
\caption{Simulation Parameters}
\label{table_1}
\begin{tabular}{|c|c|c|c|}
\hline
\textbf{Parameter} & \textbf{Value} & \textbf{Parameter} & \textbf{Value} \\ \hline
Pmin               & -20 dBm        & Pmax               & 25 dBm         \\ \hline
$N_{power}$           & 31             & Step Size          & 1.5 dBm        \\ \hline
$d_{bs_1}$               & 50 m           & $d_{fbs_1}$              & 15 m           \\ \hline
$d_{bs_2}$               & 150 m          & $d_{fbs_2}$              & 50 m           \\ \hline
$d_{bs_3}$               & 400 m          & $d_{fbs_3}$              & 125 m          \\ \hline
dth                & 25 m           &   &            \\ \hline
\end{tabular}
\end{table}
\begin{figure*}[t!]
    \centering
    \begin{subfigure}[t]{0.33\textwidth}
        \centering
        \includegraphics[width=\linewidth]{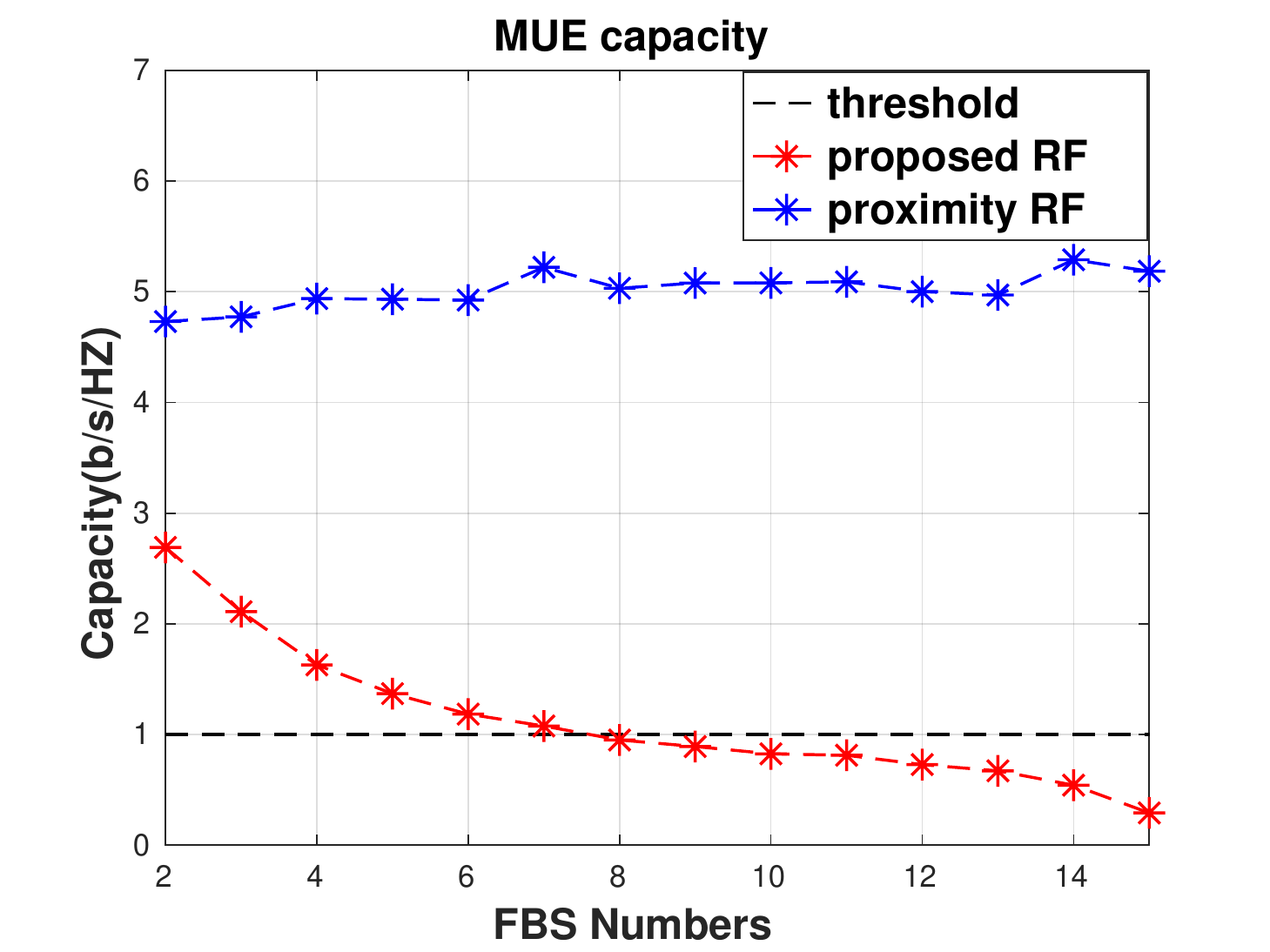}
        \caption{MUE capacity.}\label{fig:L1_MUE}
    \end{subfigure}%
    \begin{subfigure}[t]{0.33\textwidth}
        \centering
        \includegraphics[width=\linewidth]{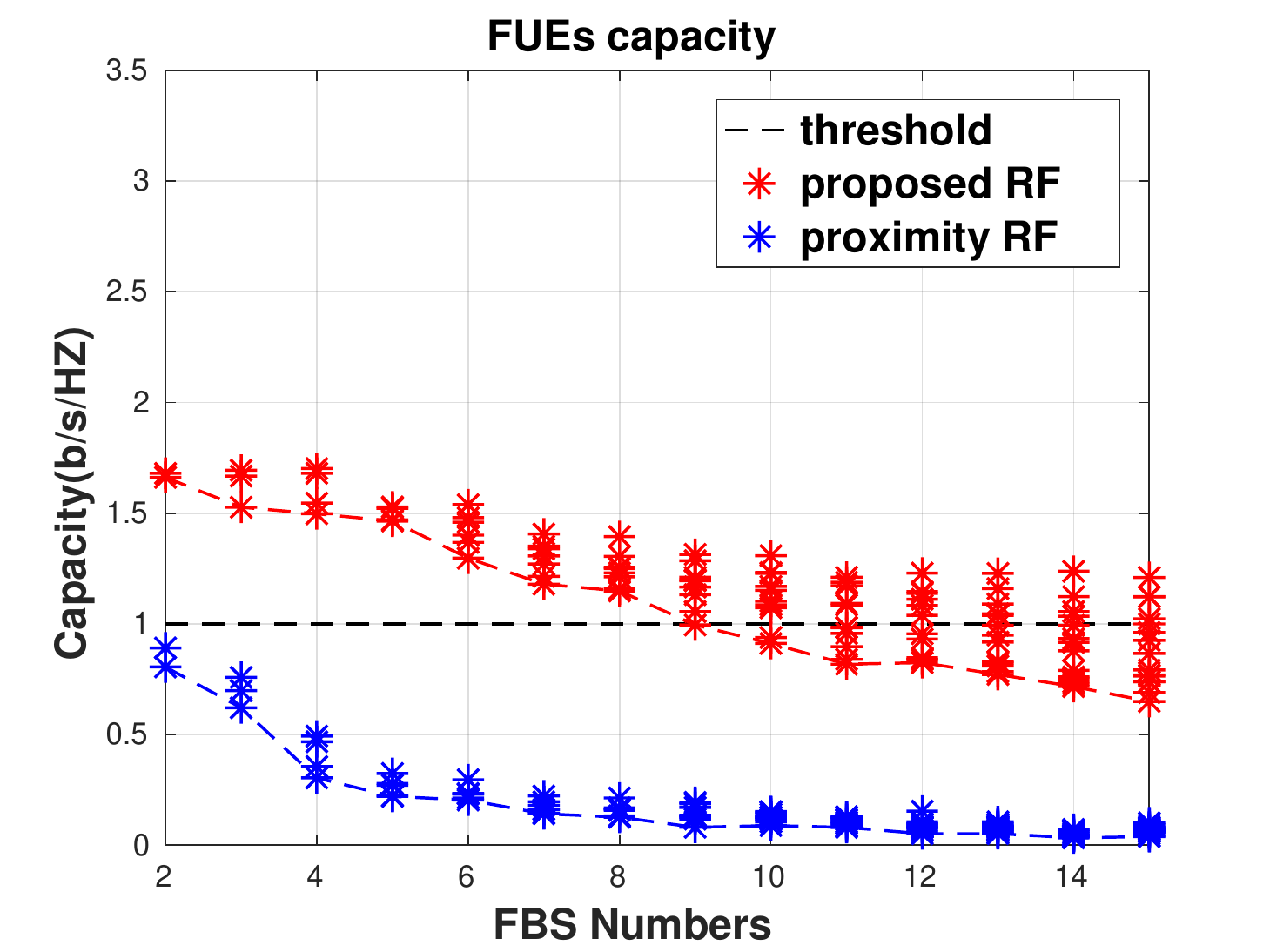}
        \caption{Capacity of FUEs.}\label{fig:L1_MINFUE}
    \end{subfigure}
    \begin{subfigure}[t]{0.33\textwidth}
        \centering
        \includegraphics[width=\linewidth]{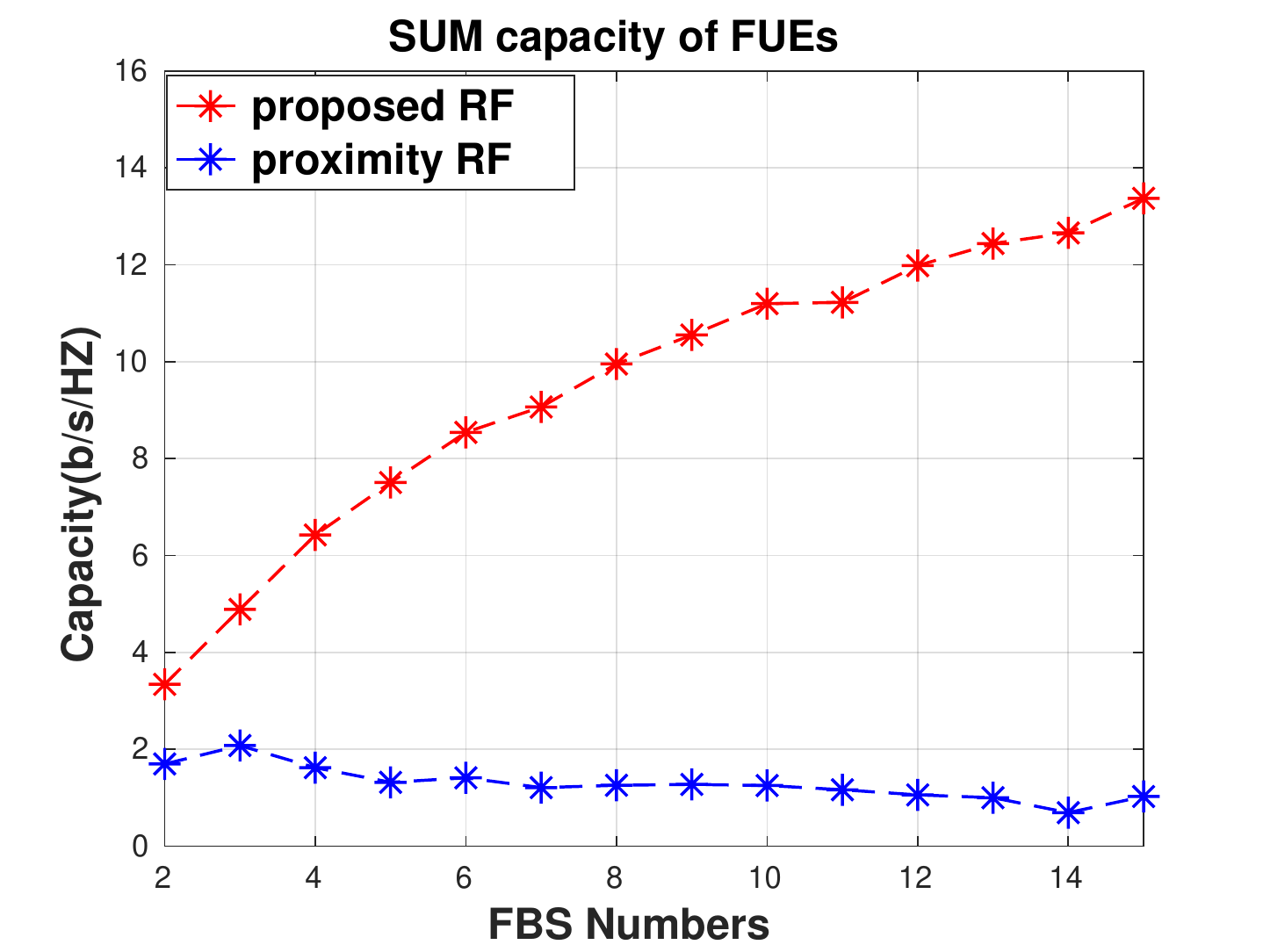}
        \caption{Sum capacity of FUEs}\label{fig:L1_SUM}
    \end{subfigure}
    \caption{Performance of the proposed reward function.}
\end{figure*}
The QoS requirements for the MUE and FUEs are defined as the capacity needed for each to support their user's application. For simulation the values of $\tilde{q}_{MUE}=1$ (b/s/Hz) and $\tilde{q}_{i}=1$ (b/s/Hz)$, i=1,..,15$ are considered for the MUE and FUEs, respectively. By knowing the MAC layer parameters, the values of the required QoS can be calculated using ~\cite[Eqs. (20) and (21)]{art_qos}. To perform Q-\textit{learning} the following values are used: learning rate $\alpha=0.5$, discount factor $\gamma=0.9$. The $\epsilon$-greedy algorithm is used for the first $80\%$ of iteration with random $\epsilon=0.1$ and the maximum number of iterations is set to $50,000$. The agents are randomly added to the network. For each number of agents, the algorithm goes through all iterations and the agents share their Q-tables according to the proposed algorithm in Section~\ref{sec_cooperation}.

\subsection{Results} \label{sec_result}
In this section we show the results of the proposed method compared to the results of the approach in~\cite{art_reward}. The method in~\cite{art_reward} is based on a proximity based reward function, which we call it \textit{proximity RF}. To have a fair comparison between the two algorithms, three measurements are plotted: the MUE capacity, the capacity of each one of the FUEs for every number of FBSs operating in the network, and the sum capacity of the FUEs. As it is shown in Fig.~\ref{fig_sim}, the position of the MUE is an example of a dense network which results in a high interference scenario. The results are shown in Figs.~\ref{fig:L1_MUE},~\ref{fig:L1_MINFUE}, and~\ref{fig:L1_SUM}, which indicate that the approach in~\cite{art_reward} is successful in satisfying the required QoS of the MUE for all number of FBSs (Fig.~\ref{fig:L1_MUE}), while it fails to support the FUEs as the density of the network increases (Fig.~\ref{fig:L1_MINFUE}). In fact, after adding the sixth FBS for the approach in~\cite{art_reward}, the FUE capacity decreases to almost zero. Hence, the QoS of the MUE is satisfied at the expense of no service for some of the FUEs. However, Fig.~\ref{fig:L1_MUE} shows that the proposed approach satisfies the QoS for the MUE and the FUEs up to the point where $8$ FBSs are operating simultaneously in close vicinity of the MUE. Further, after adding more FBSs the capacity of the MUE does not fall to zero and it is still close to the required threshold whenever $11$ FBSs are operating in close vicinity. At the same time, the majority of FUEs are still meeting their required QoS. According to Fig.~\ref{fig:L1_MINFUE} the capacity of FUEs are fairly close to each other regardless of their position, which demonstrates the algorithm's fairness.  Finally, Fig.~\ref{fig:L1_SUM} shows the sum capacity of the network which has an increasing trend for all number of FBSs and is consistently higher than that of the approach in~\cite{art_reward}.

\subsection{Convergence Analysis}

\begin{figure}[h]
\vspace{-10pt}
\begin{centering}
\includegraphics[width=0.8\columnwidth]{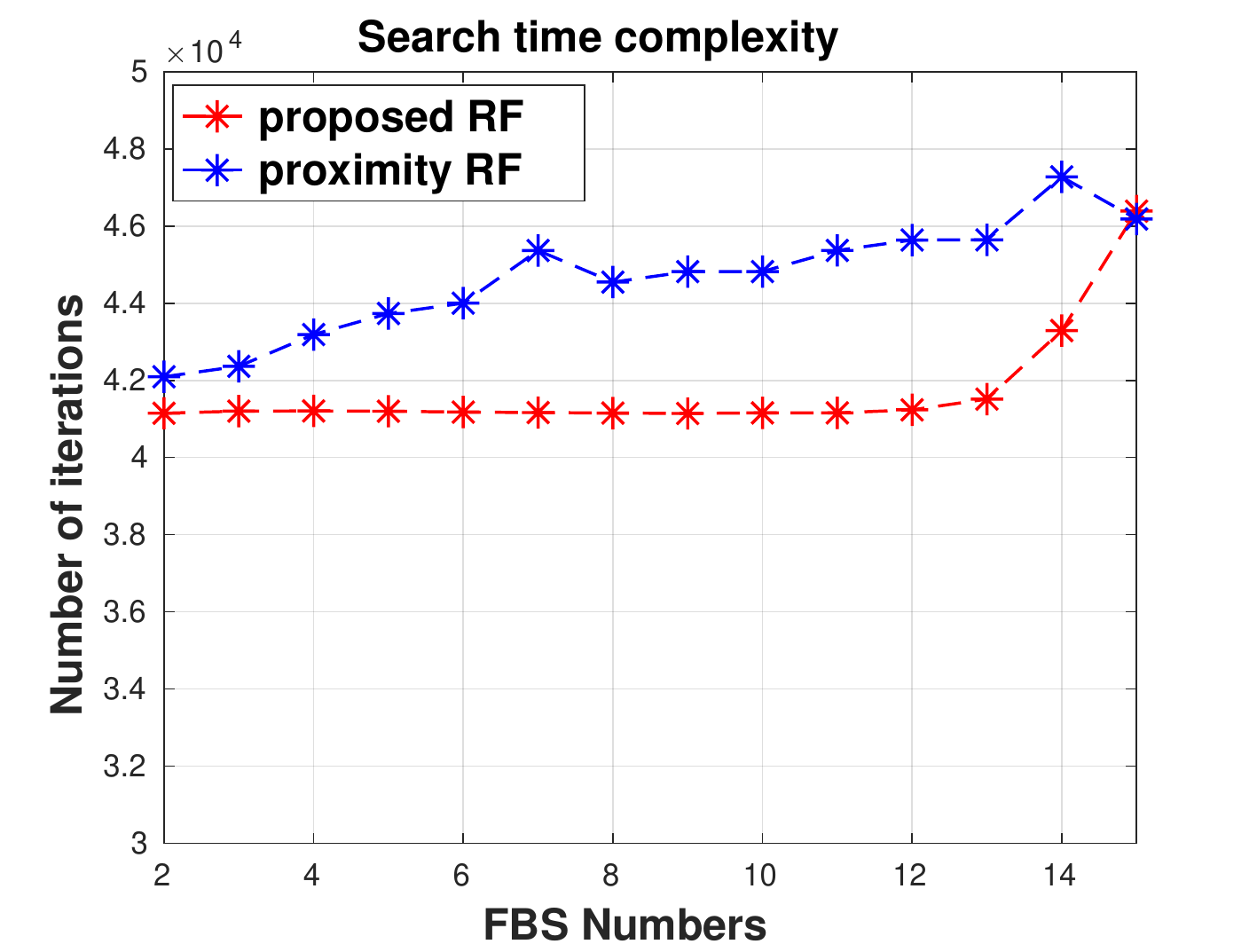}
\vspace{-5pt}
\caption[width=.3\textwidth]{Average number of iterations for the algorithms to converge.}
\label{fig_steps}
\end{centering}
\end{figure}

As it is noted in Section~\ref{sec_setup}, the maximum number of iterations to run the algorithm is set to $50,000$, although the algorithm always converges before this number is reached. Fig.~\ref{fig_steps} provides the number of iterations that it takes for both algorithms to converge with respect to the number of active FBSs in the network. As shown in Fig.~\ref{fig_steps}, the proposed algorithm requires close to $4\times10^4$ iterations whenever $13$ FBSs are active in the network. In contrast, the number of iterations is always lower than the algorithm in~\cite{art_reward}. The order of required iterations for convergence is $4\times10^4 \approx 2^{15}$, which is an extremely small portion of total number of iterations required for exhaustive search, i.e, $32^{15}= 2^{75}$.

\begin{figure}[h]
\begin{centering}
\includegraphics[width=0.8\columnwidth]{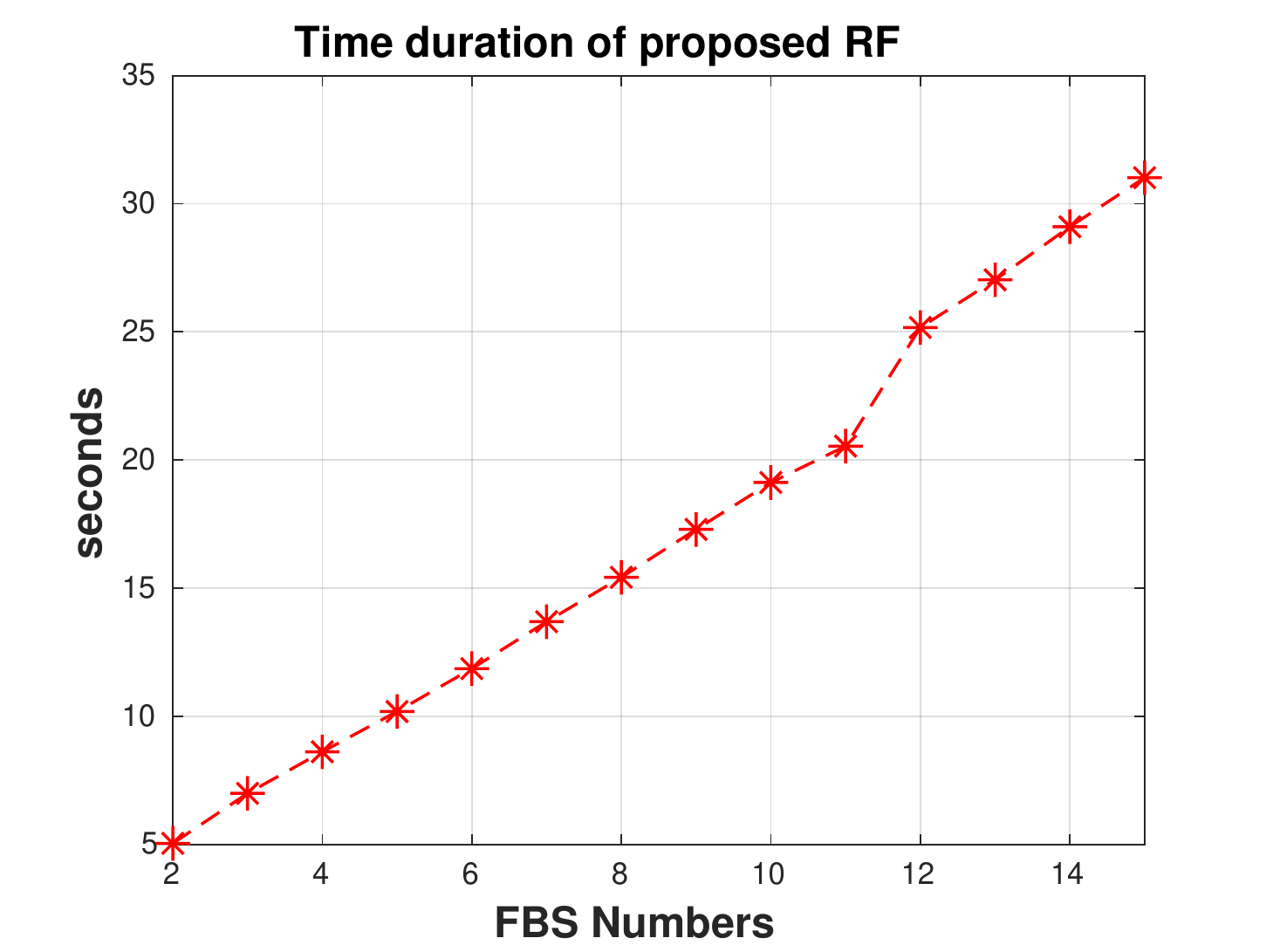}
\vspace{-5pt}
\caption[width=.3\textwidth]{Average run time of the proposed algorithm on Intel(R) Core(TM) i5-4590 CPU @ 3.30GHz.}
\label{fig_time}
\end{centering}
\end{figure}

To provide a better understanding of time duration of the proposed algorithm, Fig.~\ref{fig_time} shows the actual run time of the proposed algorithm on a regular processor.
\vspace{-20pt}

\subsection{Fairness}

\begin{figure}[h]
\vspace{-10pt}
\begin{centering}
\includegraphics[width=0.8\columnwidth]{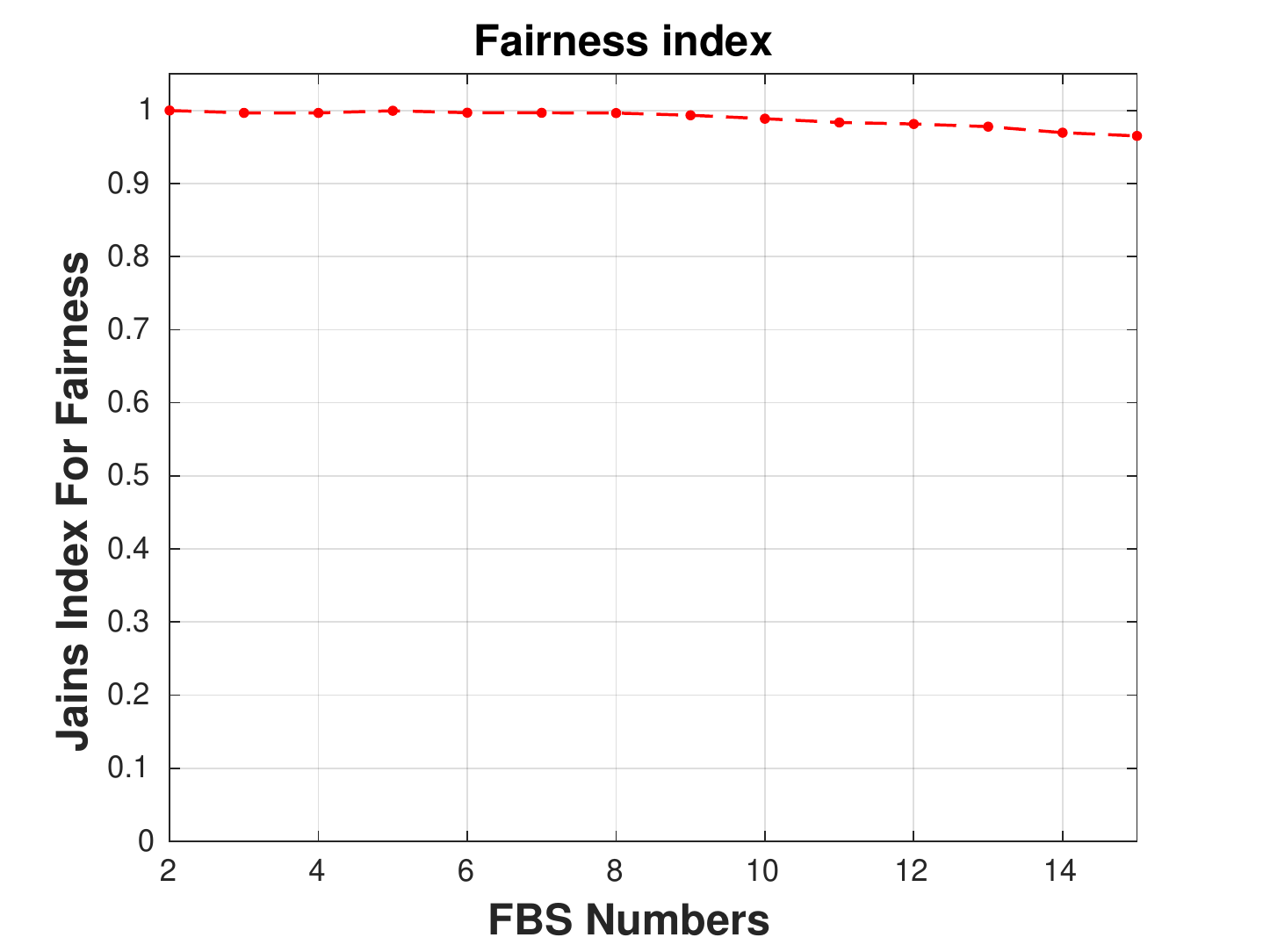}
\vspace{-5pt}
\caption[width=.3\textwidth]{Jain's fairness index as a function of the number of FBSs.}
\label{fig_fairness}
\end{centering}
\end{figure}

To provide measurement for fairness, Jain's fairness index~\cite{art_Jains} is used. In this method fairness is defined as $f(x_1,x_2,...,x_n)=\frac{\left(\sum_{i=1}^n x_i\right)^2}{n\sum_{i=1}^n x^2_i}$, in which $0 \leq f(x_1,x_2,...,x_n) \leq 1$, here equality to $1$ is achieved when all the FUEs have the same capacity. As it is shown in Fig.~\ref{fig_fairness}, the fairness index is close to one whenever $13$ FBSs are active in the network.

\subsection{Complexity Analysis}
In Section~\ref{sec_cooperation} the parameters that affect the search time of Q-\textit{learning} method are discussed. In a single FBS network (single agent network), with a finite number of iterations, and using the $\epsilon$-greedy policy with fixed $\epsilon$, and $|S|$ as the size of the state-space, the search time is upper bounded by $\mathcal{O}(|S|\log(|S|)\log(1/\epsilon)/\epsilon^2)$~\cite{art_convergence1}. The cooperation method that is proposed in Section~\ref{sec_cooperation} for femtocell networks is a special case of a learning with an external critic (LEC) method proposed by~\cite{art_complexity} for MARL networks. According to~\cite{art_complexity} the expected time needed for convergence is upper bounded by $\mathcal{O}(|S|N_{power}\log(1/\epsilon)/\epsilon^2)$, where $N_{power}$ is the number of actions (power levels) in each iteration from which the FBS can choose. $N_{power}$ is linear in state-space size. On the other hand, the optimal exhaustive search has a time complexity of $\mathcal{O}(N_{power}^M)$, where $M$ is the number of FBSs in the network.

\section{Conclusion} \label{sec_con}
The results of this paper show the application of machine learning to address resource allocation in dense HetNets. In a high interference scenarios, the power optimization in HetNet is a non-convex problem that cannot be solved with reasonable complexity. On the other hand, the proposed method as a distributed approach can solve the optimization problem in dense HetNets, while significantly reducing complexity. Our simulations show that while reducing the overall complexity of resource allocation, the proposed approach serves all users for up to $8$ femtocells whereas the approach in~\cite{art_reward} was unable to satisfy the FUEs at the expense of satisfying only the MUE. Future work will explore different methods of sharing information to obtain an optimal information sharing algorithm between agents. 


\bibliographystyle{IEEEtran}
\bibliography{IEEEabrv,powerAllocationRL}

\begin{thebibliography}{10}
\providecommand{\url}[1]{#1}
\csname url@samestyle\endcsname
\providecommand{\newblock}{\relax}
\providecommand{\bibinfo}[2]{#2}
\providecommand{\BIBentrySTDinterwordspacing}{\spaceskip=0pt\relax}
\providecommand{\BIBentryALTinterwordstretchfactor}{4}
\providecommand{\BIBentryALTinterwordspacing}{\spaceskip=\fontdimen2\font plus
\BIBentryALTinterwordstretchfactor\fontdimen3\font minus
  \fontdimen4\font\relax}
\providecommand{\BIBforeignlanguage}[2]{{%
\expandafter\ifx\csname l@#1\endcsname\relax
\typeout{** WARNING: IEEEtran.bst: No hyphenation pattern has been}%
\typeout{** loaded for the language `#1'. Using the pattern for}%
\typeout{** the default language instead.}%
\else
\language=\csname l@#1\endcsname
\fi
#2}}
\providecommand{\BIBdecl}{\relax}
\BIBdecl

\bibitem{art_Hani}
H.~Mehrpouyan, M.~Matthaiou, R.~Wang, G.~K. Karagiannidis, and Y.~Hua, ``Hybrid
  millimeter-wave systems: a novel paradigm for hetnets,'' \emph{IEEE Commun.
  Mag.}, vol.~53, no.~1, pp. 216--221, January 2015.

\bibitem{art_classic2}
Z.~Lu, T.~Bansal, and P.~Sinha, ``Achieving user-level fairness in open-access
  femtocell-based architecture,'' \emph{IEEE Trans. Mobile Comput.}, vol.~12,
  no.~10, pp. 1943--1954, Oct 2013.

\bibitem{art_ultra}
C.~Niu, Y.~Li, R.~Q. Hu, and F.~Ye, ``Fast and efficient radio resource
  allocation in dynamic ultra-dense heterogeneous networks,'' \emph{IEEE
  Access}, vol.~5, pp. 1911--1924, 2017.

\bibitem{art_macroProtect}
V.~N. Ha and L.~B. Le, ``Fair resource allocation for ofdma femtocell networks
  with macrocell protection,'' \emph{IEEE Transactions on Vehicular
  Technology}, vol.~63, no.~3, pp. 1388--1401, March 2014.

\bibitem{art_complexity}
S.~D. Whitehead, ``A complexity analysis of cooperative mechanisms in
  reinforcement learning.'' in \emph{AAAI}, 1991, pp. 607--613.

\bibitem{art_reward1}
A.~Galindo-Serrano and L.~Giupponi, ``Distributed {Q}-learning for interference
  control in {OFDMA}-based femtocell networks,'' in \emph{Proc. IEEE Veh.
  Technol. Conf.}, May 2010, pp. 1--5.

\bibitem{art_femto}
B.~Wen, Z.~Gao, L.~Huang, Y.~Tang, and H.~Cai, ``A {Q}-learning-based downlink
  resource scheduling method for capacity optimization in {LTE} femtocells,''
  in \emph{Proc. IEEE. Int. Comp. Sci. and Edu.}, Aug 2014, pp. 625--628.

\bibitem{art_reward2}
H.~Saad, A.~Mohamed, and T.~ElBatt, ``Distributed cooperative {Q}-learning for
  power allocation in cognitive femtocell networks,'' in \emph{Proc. IEEE Veh.
  Technol. Conf.}, Sept 2012, pp. 1--5.

\bibitem{art_reward}
J.~R. Tefft and N.~J. Kirsch, ``A proximity-based {Q}-learning reward function
  for femtocell networks,'' in \emph{Proc. IEEE Veh. Technol. Conf.}, Sept
  2013, pp. 1--5.

\bibitem{art_selfOptimization}
Z.~Feng, L.~Tan, W.~Li, and T.~A. Gulliver, ``Reinforcement learning based
  dynamic network self-optimization for heterogeneous networks,'' in \emph{IEEE
  Pacific Rim Conf. on Commun., Comp. and Signal Process.}, Aug 2009, pp.
  319--324.

\bibitem{art_self_fuzzy}
\BIBentryALTinterwordspacing
A.~Galindo-Serrano and L.~Giupponi, ``Self-organized femtocells: A fuzzy
  {Q}-learning approach,'' \emph{Wirel. Netw.}, vol.~20, no.~3, pp. 441--455,
  Apr. 2014. [Online]. Available:
  \url{http://dx.doi.org/10.1007/s11276-013-0609-6}
\BIBentrySTDinterwordspacing

\bibitem{art_harvesting}
M.~Miozzo, L.~Giupponi, M.~Rossi, and P.~Dini, ``Distributed {Q}-learning for
  energy harvesting heterogeneous networks,'' in \emph{Proc. IEEE. Int. Commun.
  Workshop}, June 2015, pp. 2006--2011.

\bibitem{art_cognitive}
B.~Hamdaoui, P.~Venkatraman, and M.~Guizani, ``Opportunistic exploitation of
  bandwidth resources through reinforcement learning,'' in \emph{IEEE
  GLOBECOM}, Nov 2009, pp. 1--6.

\bibitem{Watkins1992}
\BIBentryALTinterwordspacing
C.~J. C.~H. Watkins and P.~Dayan, ``{Q}-learning,'' \emph{Machine Learning},
  vol.~8, no.~3, pp. 279--292, 1992. [Online]. Available:
  \url{http://dx.doi.org/10.1007/BF00992698}
\BIBentrySTDinterwordspacing

\bibitem{book_DP}
L.~Ljungqvist and T.~J. Sargent, \emph{Recursive Macroeconomic Theory, Third
  Edition}.\hskip 1em plus 0.5em minus 0.4em\relax Cambridge, MA, USA: MIT
  Press, 2012.

\bibitem{book_sutton}
R.~S. Sutton and A.~G. Barto, \emph{Introduction to Reinforcement Learning},
  1st~ed.\hskip 1em plus 0.5em minus 0.4em\relax Cambridge, MA, USA: MIT Press,
  1998.

\bibitem{phdthesis_watkins}
C.~Watkin, ``Learning from delayed rewards,'' Ph.D. dissertation, King's
  College, Cambridge, 1989.

\bibitem{art_expertness}
M.~N. Ahmadabadi and M.~Asadpour, ``Expertness based cooperative
  {Q}-learning,'' \emph{{IEEE} Trans. Syst., Man, Cybern. {B}}, vol.~32, no.~1,
  pp. 66--76, Feb 2002.

\bibitem{art_prey}
M.~Tan, ``Multi-agent reinforcement learning: Independent vs. cooperative
  agents,'' in \emph{In Proc. ICML}.\hskip 1em plus 0.5em minus 0.4em\relax
  Morgan Kaufmann, 1993, pp. 330--337.

\bibitem{art_marl_survey}
L.~Bu¿oniu, R.~B. ${hat s}$ka, and B.~D. Schutter, ``A comprehensive survey of
  multiagent reinforcement learning,'' \emph{{IEEE} Trans. Syst., Man, Cybern.
  {C}}, vol.~38, no.~2, pp. 156--172, March 2008.

\bibitem{book_green}
E.~Hossain, V.~K. Bhargava, and G.~P. Fettweis, \emph{Green Radio Communication
  Networks}, 1st~ed.\hskip 1em plus 0.5em minus 0.4em\relax New York, NY, USA:
  Cambridge University Press, 2012.

\bibitem{pathloss}
A.~Valcarce and J.~Zhang, ``Empirical indoor-to-outdoor propagation model for
  residential areas at 0.9 -3.5 {GH}z,'' \emph{{IEEE} Antennas Wireless Propag.
  Lett.}, vol.~9, pp. 682--685, 2010.

\bibitem{art_qos}
C.~C. Zarakovitis, Q.~Ni, D.~E. Skordoulis, and M.~G. Hadjinicolaou,
  ``Power-efficient cross-layer design for {OFDMA} systems with heterogeneous
  qos, imperfect csi, and outage considerations,'' \emph{{IEEE} Trans. Veh.
  Technol.}, vol.~61, no.~2, pp. 781--798, Feb 2012.

\bibitem{art_Jains}
\BIBentryALTinterwordspacing
R.~Jain, D.~Chiu, and W.~Hawe, ``A quantitative measure of fairness and
  discrimination for resource allocation in shared computer systems,''
  \emph{CoRR}, 1998. [Online]. Available:
  \url{http://arxiv.org/abs/cs.NI/9809099}
\BIBentrySTDinterwordspacing

\bibitem{art_convergence1}
M.~J. Kearns and S.~P. Singh, ``Finite-sample convergence rates for
  {Q}-learning and indirect algorithms,'' in \emph{Advances in Neural
  Information Processing Systems 11}.\hskip 1em plus 0.5em minus 0.4em\relax
  MIT Press, 1999, pp. 996--1002.

\end{thebibliography}

\end{document}